\documentclass[onecolumn,fleqn,10pt,onecolumn]{wlscirep}
\usepackage{amsmath}
\usepackage{amssymb}
\usepackage{amsfonts}
\usepackage{graphicx}
\usepackage{hyperref}
\usepackage{url}
\usepackage{textcase}
\usepackage{bm}
\usepackage{color}
\usepackage[alsoload=hep]{siunitx}

\usepackage{ulem}

\newcommand{\Ket}[1]{\left| #1 \right\rangle} 
\newcommand{\Ketbra}[2]{\left| #1 \right\rangle \left\langle #2 \right|}

\newcommand{\piHalfPulse}{$\pi/2$-pulse~}

\newcommand{\DetTimeDep}{\Delta(t)+\dot{\Delta}(t)t} %
\newcommand{\OmegaC}{\Omega_{\textrm{C}}}
\newcommand{\OmegaZ}{\Omega_{0}}
\newcommand{\DeltaZ}{\Delta_{0}}
\newcommand{\OmegaS}{\Omega_{\textrm{S}}}
\newcommand{\DeltaS}{\Delta_{\textrm{S}}}
\newcommand{\OmegaMax}{\Omega_{\textrm{max}}}
\newcommand{\OmegaRabi}{\Omega_{\textrm{R}}}

\newcommand{\OmegaG}{\Omega}
\newcommand{\phiS}{\phi_{\textrm{S}}}
\newcommand{\tPi}{t_{\pi}}
\newcommand{\tGate}{t_{\textrm{Gate}}}

\newcommand{\HamS}{H_{\textrm{S}}}
\newcommand{\HamZ}{H_{0}}

\newcommand{\chiExp}{\chi_{\textrm{exp}}}

\newcommand{\chiZ}{\chi_{0}}
\newcommand{\Fid}[2]{F_{\textrm{#1}}^{\textrm{#2}}}
\newcommand{\FidTilde}[2]{\tilde{F}_{\textrm{#1}}^{\textrm{#2}}}

\newcommand{\SAGQG}{SAGQG}

\title{Universal, high-fidelity quantum gates based on superadiabatic, geometric phases on a solid-state spin-qubit at room temperature}

\author[1,*]{Felix Klei{\ss}ler}
\author[1]{Andrii Lazariev}
\author[1,**]{Silvia Arroyo-Camejo}
\affil[1]{Max Planck Institute for Biophysical Chemistry, Department of NanoBiophotonics, Am Fa{\ss}berg 11, 37077 G{\"o}ttingen, Germany}

\affil[*]{fkleiss@mpibpc.mpg.de}
\affil[**]{sarroyo@mpibpc.mpg.de}

\begin{abstract} 
Geometric phases \cite{1983Berry} and holonomies \cite{1984Wilczek, 1999Zanardi}  (their non-commuting generalizations) are a promising resource for the realization of high-fidelity quantum operations in noisy devices, due to their intrinsic fault-tolerance against noise and experimental imperfections.
Despite their conceptual appeal and proven fault-tolerance \cite{2009Filipp, 2012Johansson,2013Berger}, for a long time their practical use in quantum computing was limited to proof of principle demonstrations.
Only in 2012 Sjöqvist et al. \cite{2012Sjoeqvist} formulated a strategy to generate non-Abelian (i.e. holonomic) quantum gates through non-adiabatic transformation. Successful experimental demonstrations of this concept followed on various physical qubit systems \cite{2013Abdumalikov,2013Feng,2014ArroyoCamejo,Zu2014} and proved the feasibility of this fast, holonomic quantum gate concept.
Despite these successes, the experimental implementation of such non-Abelian quantum gates remains experimentally challenging since in general the emergence of a suitable holonomy requires encoding of the logical qubit within a three (or higher) level system being driven by two (or more) control fields.

A very recent proposal by Liang et al. \cite{2014Liang} offers an elegant solution generating a non-Abelian, geometric quantum gate on a simple, two-level system driven by one control field. Exploiting the concept of transitionless quantum driving \cite{2009Berry} it allows the generation of universal  geometric quantum gates through superadiabatic evolution. This concept thus generates fast and robust phase-based quantum gates on the basis of minimal experimental resources.

Here, we report on the first such implementation of a set of non-commuting single-qubit superadiabatic geometric quantum gates on the electron spin of the negatively charged nitrogen vacancy center in diamond.
The realized quantum gates combine high-fidelity and fast quantum gate performance. This provides a promising and powerful tool for large-scale quantum computing under realistic, noisy experimental conditions.
\end{abstract}

\begin{document}
\flushbottom
\maketitle

\thispagestyle{empty}

\section*{Introduction}

Currently we reside in an exciting era, in which large-scale circuit-based quantum computers do not exist yet, but their realization appears to become increasingly more feasible. This era of `Noisy Intermediate-Scale Quantum Computers' (NISQ) \cite{Preskill2018}, offers circuit-based computing platforms with $O(10)$ physical qubits and quantum annealers acting on $O(10^3)$ physical qubits. Despite these impressive achievements in scaling-up the number or qubits, a profound challenge for building viable quantum computers is yet the achievable fidelity of the fundamental quantum gates. Only when fidelity and fault-tolerance of the quantum gates are significantly improved, can quantum error correction codes be efficaciously deployed and thus universal large-scale quantum computation will become a reality. \\

Today, one of the most promising resources for intrinsically fault tolerant qubit gates are geometric (Abelian) and holonomic (non-Abelian) phases \cite{2009Filipp, 2012Johansson,2013Berger}. The quantum geometric phase was first shown to arise when a state vector is parallel-transported along a closed loop within a parameter space associated with a non-trivial state space geometry \cite{1983Berry}. The value of the geometric phase is determined by global geometric properties of the respective Hilbert space, rather than dynamic parameters. Because noise is characteristically of local nature, geometric phases are prominent to be intrinsically invariant with respect to such small control parameter imperfections. 
This intrinsic robustness of geometric phases was proposed to deliver a key performance advantage in the context of quantum computation. Zanardi and Rasetti were pioneers to propose quantum gate evolution based on holonomies, i.e. non-Abelian geometric phases \cite{1984Wilczek,1999Zanardi}. 
However, the quantum systems coherence time in combination with adiabatic system evolutions limited geometric quantum gates to proof-of-principle demonstrations without much practical relevance\cite{2000Jones}.

Only recently the generalization towards non-Abelian, non-adiabatic holonomic quantum gates (HQG) broke this limitation by nonadiabatically transporting a computational subspace in a higher dimensional Hilbert space\cite{2012Sjoeqvist}. To this end, the holonomy arises from the rotation of a complex vector (represented by a Rabi oscillation between the bright and excited states of the dressed three-level system) around a static, complex vector given by the dark state.
Experimental realizations \cite{2013Abdumalikov, 2013Feng, 2014ArroyoCamejo} of this HQG concept achieved high-fidelity quantum gate performance exceeding the threshold required for the implementation of quantum error correction protocols\cite{1995Shor,2005Knill}.
Because a holonomy can only arise in a more than two-dimensional Hilbert space the implementation of HQG requires higher-dimensional quantum systems with at least two well controlled driving fields (for examples see references \cite{2001Duan, 2012Xu, 2014Liang,2015Zhang, 2015Santos}).

In contrast, non-adiabatic geometric phases \cite{2001Wang, 2002Zhu, 2005Zhu} allow for quantum computation in a two-dimensional computational space equivalent to the systems Hilbert space at the cost of time-dependent driving fields. Until now, the realization of the latter has been pending.
Here, we report the first realization of a recently proposed single-qubit superadiabatic geometric quantum gate scheme\cite{2016Liang} which exploits the concept of transitionless quantum driving (TQD)\cite{2009Berry} to realize adiabatic state evolution in finite time (i.e., a significantly shorter time frame than conventionally suggested by the adiabatic theorem).
The gate operations are generated by controlling a single time-dependent driving field keeping the experimental resources minimal, while combining the advantages of geometric and superadiabatic evolutions.

\begin{figure}[b]
\centering
\includegraphics[scale=1]{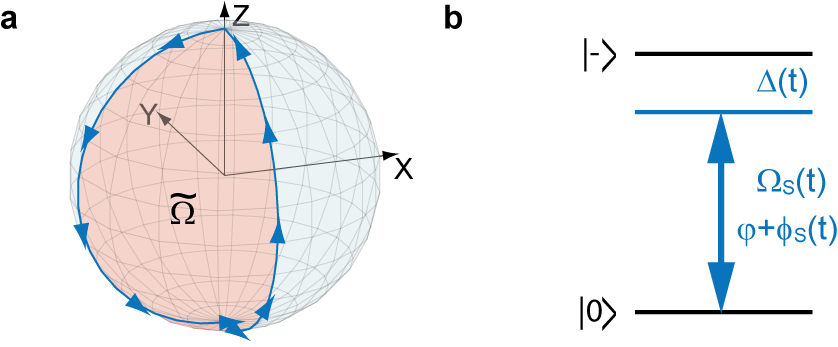}
\caption{
Superadiabtic geometric quantum gate concept. 
(a) Anticipated ``orange slice'' Bloch sphere trajectory (blue) enclosing the solid angle $\widetilde{\Omega} = 2\gamma$ (red).
(b) Two-level system and microwave field parameter (detuning $\Delta(t)$, Rabi frequency $\Omega_{\texttt{S}}(t)$ and phase $\varphi+\phiS(t)$) utilized for the realization of superadiabatic geometric quantum computation.}
\label{fig:general}
\end{figure}
Experiments are performed utilizing the electron spin dedicated to the nitrogen vacancy (NV) center in diamond, a promising candidate for the implementation of a scalable quantum registers.
Dynamic single-qubit\cite{2002Kennedy} and multi-qubit \cite{2012Dolde} gates as well as non-adiabatic non-Abelian geometric single-qubit gates\cite{2014ArroyoCamejo} have demonstrated its significance for quantum information applications, even at room-temperature.
Moreover, the use of optimized samples eliminates/supresses the noise environment as source of error and high fidelity quantum computation can be obtained by choosing quantum operations insensitive to control parameter imperfections.

\section*{Results}

\subsection*{Superadiabatic geometric quantum gates}

The superadiabatic geometric quantum gate (SAGQG) proposal \cite{2016Liang} builds upon the concept of the Aharonov-Anandan type non-adiabatic geometric phase \cite{1987Aharonov}. For the Aharonov-Anandan phase to be solely of geometric nature, in the total phase 
\begin{equation}
\Phi = -\int_0^T \langle \psi (t)| H(t)|\psi(t)\rangle \textrm{d}t + \int_0^T \langle \widetilde{\psi} (t)| i (\partial/\partial t) \widetilde{\psi}(t)\rangle \textrm{d}t
\end{equation}
the first, dynamic phase term must vanish (here $|\widetilde{\psi}(t)\rangle$ is the reference section state on the projective Hilbert space $\mathcal{P}$ \cite{1995Pati}). This can be achieved by driving the state vector with a driving field that is applied perpendicularly to the state vector at all times. Under this condition driving the state vector on the Bloch sphere, the solid angle $\widetilde{\Omega}$ enclosed by the Bloch vector trajectory determines the acquired geometric phase $\gamma = \Phi = \widetilde{\Omega}/2 $ (Fig.~\ref{fig:general}a).

The Aharonov-Anandan phase is restricted to generate U(1) phase shift gates. The total Hamiltonian of the SAGQG is constructed employing the technique of transitionless driving \cite{2009Berry} where a reverse engineered correction Hamiltonian compensates for undesired transitions between the basis states. This way the effective superadiabatic Hamiltonian drives the instantaneous eigenstates exactly such that the evolution of dynamic phases is fully suppressed, even within the fast driving regime. An alternative to this special constraint is the simple cancellation of the arising dynamic phase by designing self-compensating trajectories.

Considering a two-level system with time-dependent single driving field our original Hamiltonian $\HamZ(t)$ following form in the co-rotating reference frame of the external driving field
\begin{eqnarray}
\label{eq:Hamiltonian0}
\HamZ=\frac{\hbar}{2}
\begin{pmatrix}
\Delta(t)+\dot{\Delta}(t)t & \OmegaRabi(t) e^{- i \varphi}\\
\OmegaRabi(t) e^{i \varphi} & -(\Delta(t)+\dot{\Delta}(t)t)
\end{pmatrix} \ ,
\end{eqnarray}
where the driving field is applied with a detuning $\Delta(t)$, phase $\varphi$, and Rabi frequency $\OmegaRabi(t)$. The non-standard form of the Hamiltonian in Eq.~\eqref{eq:Hamiltonian0} in the rotating frame of the driving field arises from its time-dependent detuning (see Supplementary Information for details on the derivation of Eq.~\eqref{eq:Hamiltonian0}).
Exploiting the concept of transitionless quantum driving \cite{2009Berry} and deriving a suitable correction Hamiltonian $H_{\textrm{c}}$ Liang et al. \cite{2016Liang} propose the superadiabatic Hamiltonian  
\begin{eqnarray}
\HamS(t) = H_{0} + H_{\textrm{c}} = \frac{\hbar}{2}
\begin{pmatrix}
\Delta(t)+\dot{\Delta}(t)t & \OmegaS(t) e^{- i [\varphi + \phiS(t)]}\\
\OmegaS(t)e^{i [\varphi + \phiS(t)]} & -(\Delta(t)+\dot{\Delta}(t)t)
\end{pmatrix}\ ,
\label{eq:SAGHamiltonian}
\end{eqnarray}
where $\OmegaS(t) = \sqrt{\OmegaRabi(t)^2+\OmegaC(t)^2} $ is the superadiabatic Rabi frequency, and $\phiS(t) = \arctan\left[\OmegaC(t)/\OmegaRabi(t)\right]$ is the superadiabatic phase.
The corrected Rabi frequency is $\OmegaC(t) = \left[\dot{\Omega}_{\textrm{R}}(t)(\DetTimeDep)-\OmegaRabi(t)\partial_{t}(\DetTimeDep)\right]/\OmegaG^2$, where the generalized Rabi frequency is introduced as $\OmegaG = \sqrt{\OmegaRabi(t)^2+(\DetTimeDep)^2}$.
The instantaneous eigenstates of the superadiabatic Hamiltonian $\HamS(t)$ are $\Ket{\lambda_{\pm}(t)}$. (The explicit expression of the superadiabatic Rabi frequency $\OmegaS(t)$, detuning $\DeltaS(t) = \DetTimeDep$ and phase $\varphi(t)$ are given in the methods part.)

In order to realize universal quantum computation the SAGQG applies a strategy previously developed by Wang and Zhu \cite{2002Zhu, 2003bZhu} which is based on choosing a pair of orthogonal states $\Ket{\lambda_{\pm}(t)}$ undergoing a cyclic evolution: $\Ket{\lambda_{\pm}(T)}=\exp\left[i \phi_{\pm}\right]\Ket{\lambda_{\pm}(0)}$.
Over the full length of the SAGQG transformation the dynamic phase is designed to cancel such that the system evolution becomes fully geometric $U(T,0) \Ket{\lambda_{\pm}(t)} = \exp\left[\pm i \gamma\right]\Ket{\lambda_{\pm}(0)}$ where the evolution operator $U(T,0)$ imprints only a U(1) phase factor on each of the eigenstates $\Ket{\lambda_{\pm}(0)}$.
Wang and Zhu ingeniously identified that these trivial phase factors on the $\Ket{\lambda_{\pm}(0)}$ nevertheless translate to a non-Abelian transformation on the computational states in the co-rotating frame. 
Thus, even though the SAGQG is not based on a non-Abelian holonomy, in virtue of the elaborate basis transformation between the cyclic states and the computational states the U(1) geometric phase factors translate to a non-Abelian, geometric transformation of the computational states allowing for universal quantum computation.

\subsection*{Bloch sphere trajectory}

\begin{figure}[tb]
\centering
\includegraphics[scale=1]{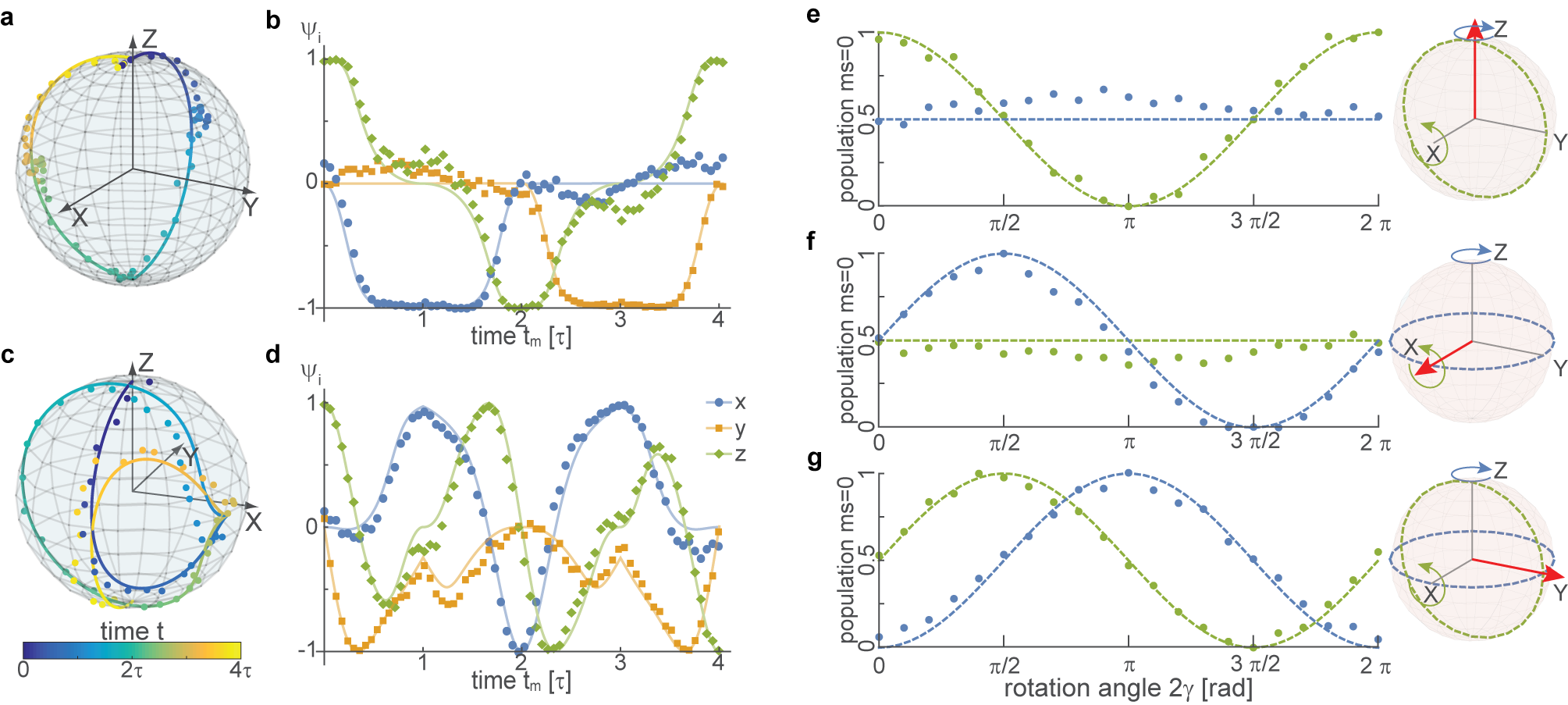}
\caption{Superadiabatic geometric gate realization: (a) Simulated and reconstructed Bloch sphere trajectory of the superadiabatic geometric Pauli-Z gate in the driving field frame for a spin initialized into the $m_{s}=0$ state.
(b) Bloch vector component $\Psi_{x}$ (blue), $\Psi_{y}$ (orange) and $\Psi_{z}$(green) of the trajectory presented in (a) versus the gate time in multiples of $\tau$.
Solid lines represent numerically calculated trajectories and dots indicated measured values.
Analogously (c) and (d) follow for the realized Pauli-X gate.
(e-g) Measured population of the $\Ket{0}$ state for a spin initialized into the orthogonal states (e) $\Ket{0}$, (f) $1/\sqrt{2} (\Ket{0}-\Ket{1})$ and (g) $1/\sqrt{2}(\Ket{0}+i\Ket{1})$ in dependence of $\gamma$ for superadiabatic rotations around the x (green) and z-axis (blue).
Dashed lines represent the expected values.
Bloch spheres indicate the initialized state (red arrow).
}
\label{fig:Figure2}
\end{figure}

The SAGQG state evolution is based on a sequence of four trajectory segments of time duration $\tau$, leading to a total gate length of $t_{\mathrm{Gate}}=4\tau$ (see Methods section for details).
We investigate and visualize the quantum gate Bloch sphere trajectory of a qubit initialized into the $m_{s} = 0$ state in a stroboscopic manner by applying projective readout pulses at times $t_{m}$.
As two representative gates we realize the Pauli-Z (Fig.~\ref{fig:Figure2}a, b) and the Pauli-X (Fig.~\ref{fig:Figure2}c, d) gate.
The measured Bloch vector trajectories (dots) are in very good agreement with the  numerically calculated trajectories (solid lines).
Rotations around the y-axis (Pauli-Y gate) can be realized by setting $\varphi = \pi/2$ for the original Hamiltonian $\HamZ(t)$.
In the realization of the Pauli-Z gate the, non-adiabatically obtained original trajectory is observed (compare Fig.~\ref{fig:general}a).
The particular shape of the trajectory in Fig.~\ref{fig:Figure2}c, d illustrates that the short-cut to adiabaticity is obtained utilizing a sophisticated parameter time-dependence.

\subsection*{Generalization to geometric phase gate with arbitrary value}

So far we demonstrated that rotations by $\gamma = \pi/2$ around the x and z-axis can be fulfilled with high fidelity by performing superadiabatic geometric quantum computation.
In addition, by varying the opening angle of the ``orange slice'' trajectory an arbitrary geometric phase $\gamma$ can be acquired.
Utilizing the orthogonal states $\Ket{0}$, $1/\sqrt{2}(\Ket{0}-\Ket{1})$ and $1/\sqrt{2}(\Ket{0}+i \Ket{1})$ we demonstrate the rotation for different geometric phases $\gamma$ (see Fig.~\ref{fig:Figure2}e-f).
In order to visualize the phase gate we map the acquired phase into a population by application of a projective \piHalfPulse around the $\bar{y}$-axis.
Hence, we show that the SAGQG concept additionally allows for the generation of an arbitrary phase shift gate. Collectively with the former we thus provide a universal set of single-qubit geometric quantum gates.

\begin{table}[tb]
\caption{Experimentally obtained corrected quantum gate fidelities $\tilde{F}$ and average gate error $\epsilon_{g}$ of the single-qubit SAGQGs.}
\label{tab:Fidelities}
\centering
\begin{tabular}{c|c|c||c}
$\tilde{F}_{X}$ 			& $\tilde{F}_{Z}$ 			& $\tilde{F}_{H}$ 			& $\epsilon_{g}$\\ \hline
$0.994_{-0.031}^{+0.026}$	& $0.995_{-0.024}^{+0.021}$	& $0.992_{-0.029}^{+0.022}$	&  $0.0013 (3)$	
\end{tabular}
\end{table}

\subsection*{Fidelity assessment and fault-tolerance}

Quantification of the performance of the superadiabatic geometric gates is obtained via standard quantum process tomography (QPT)\cite{2004OBrien} measurements, which allows to reconstruct the full experimental quantum process matrix $\chiExp$ and therefore to determine the quantum gate fidelity $F=\textrm{Tr}\left(\chiExp\chiZ\right)$\cite{NielsenChuang}, where $\chiZ$ is the theoretically anticipated process matrix (for details on the experimental QPT procedure see Supplementary Information and reference \cite{2014ArroyoCamejo}).
Due to their dynamic nature and finite time duration the QPT pulses are susceptible to errors and we obtain the corrected quantum gate fidelity value $\FidTilde{}{}=F/\Fid{ID}{}$ by normalization with the fidelity of the identity operation.
We determine the experimental gate fidelities of the SAGQG to be $\FidTilde{x}{\SAGQG}=0.994_{-0.031}^{+0.026}$ and $\FidTilde{z}{\SAGQG}=0.995_{-0.024}^{+0.021}$ for Pauli-X and Pauli-Z operations, respectively. 
Additionally, the Hadamard gate is realized by a rotation of $\pi/2$ around the y-axis ($R_{y}(\pi/2)$) and a subsequent rotation by $\pi$ around the z-axis ($R_{z}(\pi)$), resulting in an experimental fidelity of $\FidTilde{H}{\SAGQG}=0.992_{-0.029}^{+0.022}$.
These values clearly exceed the necessary fidelity threshold on the order of $1-10^{-2}$ for the implementation of state-of-the-art error correction codes based on, e.g., surface codes \cite{1997Kitaev,2012Fowler}. The SAGQG concept thus qualifies as a promising candidate for the implementation of scalable quantum computing.\\

\begin{figure}[tb]
\centering
\includegraphics[scale=1]{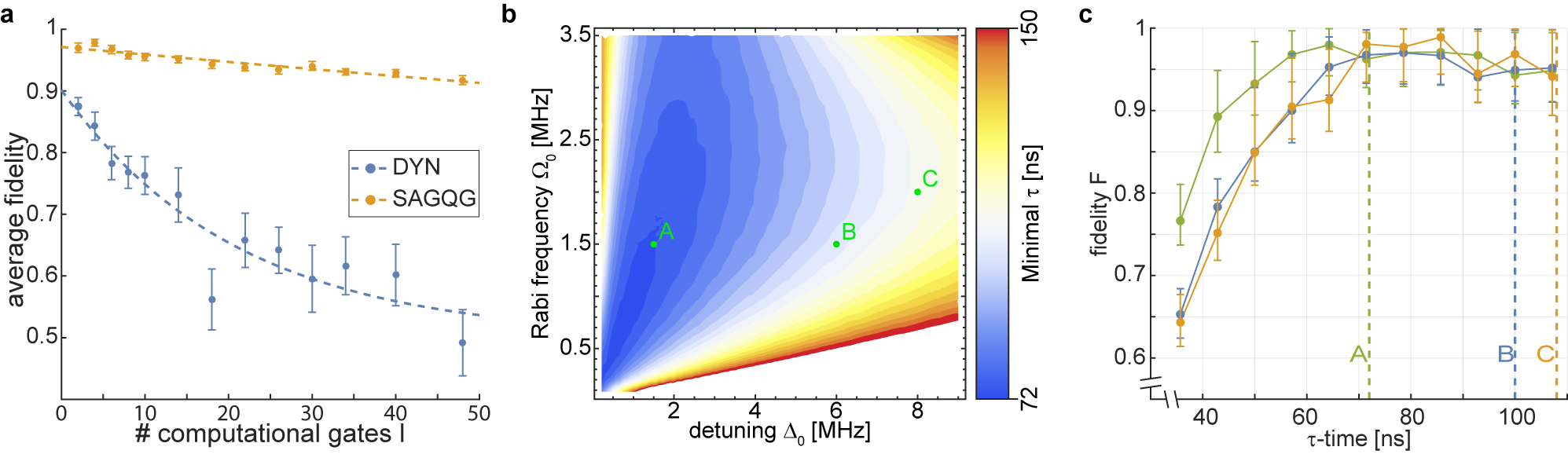}
\caption{Robustness analysis: 
(a) The randomized benchmarking analysis reveals the decay of the average fidelity in dependence of the number of computational gates $l$ for a set of SAGQG (orange) and a set of dynamic quantum gates (blue).
The average probability of error per gate are $\epsilon_{g}^{\mathrm{SAGQG}}=0.0013(3)$ and $\epsilon_{g}^{\mathrm{Dynamic}}=0.023(8)$, respectively.
Error bars represent the standard error of the mean.
(b) Minimal value of $\tau$ in dependence of the free parameter $\OmegaZ$ and $\DeltaZ$ for a system with maximal Rabi frequency $\OmegaMax = \SI{7}{\MHz}$.
(c) Measured quantum gate fidelity $\Fid{}{}$ as a function of $\tau$ for three free parameter combinations indicated in (b) by A, B and C.
Solid lines are a guide to the eye.
Vertical dashed lines represent the numerically calculated minimal $\tau$ value fulfilling $\OmegaS(t,\tau,\OmegaZ,\DeltaZ) \leq \OmegaMax$.
}
\label{fig:Figure3}
\end{figure}

Besides the fidelity of the individual, logical gates, we additionally assess the  average error probability over the set of universal gates employing randomized benchmarking\cite{2008Knill}. Based on the application of randomly assembled sequences of a set of logical gates, randomized benchmarking allows for a good estimation of the error scaling given a long sequence of quantum gates, as relevant for viable applications in longer quantum algorithms. Fig.~\ref{fig:Figure3}a presents the average fidelity as a function of the number of computational gates $l$. For the SAGQG we obtain an average probability of error per gate of $\epsilon_{g}^{\mathrm{SAGQG}} = 0.0013 (3)$, whereas an identical analysis for a set of dynamic quantum gates represented by $\pi$ and $\pi/2$-pulses  reveals an average probability of error of $\epsilon_{g}^{\mathrm{dynamic}} = 0.023 (8)$, i.e. the geometric-phase based SAGQG performs one order of magnitude better than its dynamic-phase based standard gate (see Supplementary Information for details). Our results suggest that the SAGQG is significantly more resilient with respect to the type of noise and parameter imperfections present in our experimental system than the standard realization of dynamic phase-based quantum gates. This experimental finding strongly supports the long time conjectured robustness \cite{1999Zanardi} of geometric phase-based quantum gates, which owes to the fact that geometric phases and holonomies are global features which are intrinsically robust with respect to locally occurring parameter imperfections and noise that leave the state-space area enclosed by the states trajectory on the respective projective space invariant. \\

In the following we examine the fidelity performance of the SAGQG with respect to variations in the gate evolution time. 
This is important for two reasons: 
1) In order to most efficiently exploit the coherence time of the qubit, we need to investigate the theoretical velocity limits and experimental performance of the SAGQG and aim for fast quantum gate performance. 
2) We experimentally examine the fault-tolerance of the SAGQG with respect to experimental parameter imperfections. In particular we analyse the SAGQG performance outside its optimal parameter specifications. The latter is particularly relevant for the common experimental case where the Rabi frequency (for practical reasons) obeys a maximum bound $\max_t \left(\OmegaS(t, \Omega_0, \Delta_0)\right) \leq \OmegaMax$ (for parameter dependences see Methods section). Given such a practical maximum bound  $\OmegaMax$ for the experimentally achievable Rabi frequency, in Fig.~\ref{fig:Figure3}b we show a contour plot of the numerically determined minimal $\tau$-value, denoted $\tau_{\textrm{min}}(\Omega_0, \Delta_0)$, fulfilling the necessary criterion $\max_t \left(\OmegaS(t, \Omega_0, \Delta_0)\right) \leq \OmegaMax$.
We like to stress again, the $\tau_{\textrm{min}}(\Omega_0, \Delta_0)$ limit is not given by theoretical constraints related to the state evolution (e.g., adiabaticity), but it is merely defined by the experimentally achievable Rabi strength $\OmegaMax$. The smallest, experimentally feasible $\tau$-value is equivalent to $1/(2\,\OmegaMax)$ corresponding to the length of a $\pi$-pulse $\tPi$, ultimately limiting the SAGQG length to $\tGate \geq 2/\OmegaMax = 4\tPi$. For our experimental conditions the minimal gate length $t_{\textrm{Gate}}=4\tau$ corresponds to $t_{\mathrm{Gate}}=\SI{284}{\nano\second}$. If $\tau$ were chosen smaller than $\tau_{\textrm{min}}$ this would require $\max_t \left(\OmegaS(t, \Omega_0, \Delta_0)\right) $ to exceed $\OmegaMax$ which -- given experimental limitations on $\OmegaMax$ -- cannot be fulfilled by any experimental parameter set. Forcing $\tau<\tau_{\textrm{min}}$ experimentally leads to a marked mismatch between required and actual value of the driving field strength $\OmegaS(t)$, i.e. an inconsistent, erroneous driving field parameter set.

For an experimental robustness analysis of the SAGQG we explicitly vary the gate time parameter $\tau$ within a non-optimal range of $\tau$ reaching from $0.5\cdot \tPi$ to $1.5\cdot \tPi$ (whereas the theoretical  $\min_{\Omega_0, \Delta_0}\left(\tau_{\mathrm{min}}\right)=\tPi$) for three sets of parameters A, B, and C ($\OmegaZ= \{1.5, 1.5, 2\}$~MHz and $\DeltaZ = \{1.5, 6, 8\}$~MHz). 
The $\min_{\Omega_0, \Delta_0}\left(\tau_{\mathrm{min}}\right)$ value for each parameter set is marked in Fig.~\ref{fig:Figure3}c as a vertical, dashed line of matching colors, respectively.
Fig. \ref{fig:Figure3}c shows the extracted quantum gate fidelity $F$ of the Pauli-X gate in dependence of $\tau$. We observe that even for $\tau$ smaller than the calculated threshold $\tau_{\mathrm{min}}$ (indicated by vertical dashed lines) the quantum gate fidelity $F$ remains close to one. Only for $\tau < \tPi \approx \SI{71}{\nano\second}$ is the fidelity dropping. These results proof the tolerance of the SAGQG to perform stably over a large range of timing parameter variations and give evidence for the intrinsic robustness of the SAGQG against timing imprecision and concomitant mismatches in the driving field strength.

\section*{Discussion}

In this work we demonstrated for the first time the experimental realization of the recently proposed universal set of single-qubit superadiabatic geometric quantum gates, utilizing the NV\,center electron spin in diamond at room temperature. Our experimental demonstration exhibits fast and high-fidelity qubit gate performance while requiring only a minimalistic qubit and control system for its realization, if compared to schemes based on holonomic qubit gates reaching similar high-fidelity performance. The realization within a two-level system sets comparatively low requirements on the experimental apparatus and the single driving field reduces the number of control parameters significantly. 
We explicitly investigated and confirmed the tolerance of this gate type with respect to errors in the gate time and experimentally verified its robustness. 

An extension of the SAGQG concept to a two-qubit controlled-NOT and controlled-PHASE has been proposed \cite{2016Liang} and would, together with the single-qubit set presented here, provide a universal set of superadiabatic geometric quantum gates. Beyond the demonstration in this work performed on an NV center spin qubit, this single-qubit gate technique is directly translatable to other promising experimental qubit systems, like, e.g., atomic, ion, transmon or flux qubits. Beyond quantum computing, the SAGQG concept presented here could be employed as a universal, high-fidelity building block for other novel quantum technologies being fundamentally based on quantum operations, like quantum communication or qubit-assisted nanosensing applications.

\section*{Methods}

\subsection*{Original Hamiltonian}
\label{sec:DrivingParameters}
In the following the driving field parameter for the realization of superadiabatic phase gate according to Liang \textit{et al.}\cite{2016Liang} are listed.
The Rabi frequency used to yield $\OmegaS(t)$ reads
\begin{eqnarray}
\OmegaRabi(t) = 
\left\lbrace
\begin{matrix}
\OmegaZ \left[1-\cos\frac{\pi t}{\tau} \right], & 0\leq t < \tau\\ \vspace{0.05 cm}
\OmegaZ \left[1+\cos\frac{\pi (t - \tau)}{\tau} \right],& \tau \leq t  < 2 \tau\\ \vspace{0.05 cm}
\OmegaZ \left[1-\cos\frac{\pi (t - 2\tau)}{\tau} \right], & 2 \tau \leq t < 3 \tau\\ \vspace{0.05 cm}
\OmegaZ \left[1+\cos\frac{\pi (t - 3\tau)}{\tau} \right], & 3 \tau \leq t \leq 4 \tau 
\end{matrix}
\right.
\end{eqnarray}
To obtain the experimentally relevant detuning $\Delta(t)$ of the driving field the following differential equation needs to be solved for:
\begin{eqnarray}
\Delta_{\textrm{S}}(t)=\DetTimeDep = 
\left\lbrace
\begin{matrix}
\DeltaZ \left[\cos\frac{\pi t}{\tau} +1 \right], & 0\leq t < \tau\\ \vspace{0.05 cm}
\DeltaZ \left[\cos\frac{\pi (t - \tau)}{\tau} -1 \right],& \tau \leq t  < 2 \tau\\ \vspace{0.05 cm}
\DeltaZ \left[\cos\frac{\pi (t - 2\tau)}{\tau} +1 \right], & 2 \tau \leq t < 3 \tau\\ \vspace{0.05 cm}
\DeltaZ \left[\cos\frac{\pi (t - 3\tau)}{\tau} -1 \right], & 3 \tau \leq t \leq 4 \tau 
\end{matrix}
\right.
\end{eqnarray}
The value of the acquired geometric phase $\gamma = \pi - (\widetilde{\varphi}_{1}-\widetilde{\varphi}_{2})$ is defined by constant phases $\widetilde{\varphi}_{1}$ and $\widetilde{\varphi}_{2}$ added to the driving field phase 
\begin{eqnarray}
\varphi +\phi_{s}(t)
=
\left\lbrace
\begin{matrix}
\widetilde{\varphi}_{1} + \phi_{s}(t), & 0 \leq t < 2\tau \\
\widetilde{\varphi}_{2} + \phi_{s}(t), & 2\tau \leq t \leq 4\tau
\end{matrix}
\right.
.
\end{eqnarray}
For the realization of the Pauli-Z gate presented here we set $\widetilde{\varphi}_{1}=0$ and $\widetilde{\varphi}_{2}=\pi/2$ resulting in the wanted phase value of $\gamma=\pi/2$.
The driving field parameter for the realization of a spin-flip gate follow in a similar manner and are shown in the supplementary material explicitly.

\subsection*{Nitrogen vacancy center in diamond}

The NV center consists of a substitutional nitrogen atom and an adjacent vacant lattice site in the carbon diamond lattice.
A spin-one system is associated with the negatively charged NV species, which can be efficiently initialized\cite{2006Gaebel} and readout\cite{2007Dutt} by optical means.
The triplet ground state features a zero field splitting of $D \approx 2\pi \times \SI{2.87}{\GHz}$ between the $\Ket{0}$ and $\Ket{-},\Ket{+}$ states.
Aligning an external magnetic field of $\left|B\right| \approx \SI{400}{\gauss}$ along the NV center axis enables dynamic nuclear polarization of the nitrogen nuclear spin\cite{2009Smeltzer,2009Jasques} and sets the triplet transition frequencies to $\Ket{0}\leftrightarrow \Ket{-}(\omega_{0-} \approx 2\pi \times \SI{1.73}{\GHz})$ and $\Ket{0} \leftrightarrow \Ket{+}(\omega_{0+} \approx 2\pi \times \SI{4.01}{\GHz})$.
Both transitions can be manipulated coherently by applying microwave fields at frequencies $\omega_{-}=\omega_{0-}+\delta_{-}$ and $\omega_{+}=\omega_{0+}+\delta_{+}$, where $\delta_{\pm}$ is the detuning from the resonance.
For our experiments we employ the two-level systems comprised of the $|0\rangle$ and $|-\rangle$ states, as illustrated in Fig.~\ref{fig:general}b.

\subsection*{Experimental realization}

A custom-made confocal microscope equipped with a $546$~nm cw laser serves for optical initialization of the NV spin qubit and facilitates optical readout of the final spin states from the NV spin's emitted fluorescence intensity. Coherent microwave manipulation is conducted by means of an arbitrary waveform generator (AWG) that can be programmed at a high sampling rate  of $25\,\textrm{GSamples/s}$ as needed. While the Rabi frequency and detuning of the applied MW field needed to follow specific time-dependences, maximum values were $\OmegaS = 7$~MHz and $\Delta_{\textrm{S}}=2$~MHz.

The employed NV center was generated in an isotopically pure diamond from Element 6 ($99.999\textrm{\,\%}$ $^{12}\textrm{C}$ abundance) as grown diamond substrate, by $^{14}\textrm{N}$ ion implantation at around $10$~MeV, leading to the formation of NV center in a depth of around $\SI{3.7}{\micro\metre}$  below the diamond surface after annealing.
We determine a longitudinal relaxation time of $T_{1} = (5\pm \SI{0.5})~{\milli\second}$ and a spin-dephasing time of $T_{2}^{*} = (4.25 \pm \SI{0.27})~{\micro\second}$.
At a magnetic field of $\approx\SI{402}{\gauss}$ aligned along the NV center axis we obtain a nuclear polarization of $0.94 \pm 0.05$ into the $m_{I}=+1$ hyperfine state.

\section*{Acknowledgments }
The authors acknowledge funding by the VW-Stiftung.
We thank Element 6 for providing the diamond sample in the framework of the DARPA QUASAR project. We thank Junichi Isoya for the nitrogen implantation and Philipp Neumann for annealing of the diamond sample. We want to thank Erik S\"{o}qvist for helpful comments on the manuscript. We thank Stefan W. Hell for his support throughout the project.

\section*{Author contributions statement}
F.K., S.A.-C. and A.L. performed and evaluated the experiments.
S.A.-C. and F.K. performed analytical and numerical calculations and data analysis.
A.L. and S.A.-C. designed and programmed the experiment control software.
F.K. and S.A.-C. wrote the manuscript.
S.A.-C. conceived and supervised the project.

\section*{Additional information}
\subsection*{Competing financial interests:}
The authors declare no competing financial interests.


\newpage

\begin{center}
\LARGE{\textbf{Supplementary Information:}}\\
\end{center}

\section{Accelerated driving field frames}
\label{sec:AcceleratedDrivingFieldFrames}
The superadiabatic geometric quantum gate is defined on the two-dimensional Hilbert space $\mathcal{H}^2$ and is driven by a time-dependent field.
Consequentially, the Hamiltonian in the laboratory frame $H_{L}(t)$ follows as the one of a two-level system with energy spacing $\hbar \omega_{0}$ and a driving field of angular frequency $\omega_{D}(t)$, phase $\varphi$ and amplitude $\OmegaRabi(t)$:
\begin{eqnarray}
H_{L}(t) &=& 
\frac{\hbar}{2}
\begin{pmatrix}
\omega_{0} & 2\OmegaRabi(t) \cos(\omega_{D}(t)t+\varphi) \\
2\OmegaRabi(t) \cos(\omega_{D}(t)t+\varphi) & -\omega_{0}
\end{pmatrix} \nonumber
\\
&=&
\frac{\hbar}{2}
\begin{pmatrix}
\omega_{0} & \OmegaRabi(t) (e^{i (\omega_{D}(t)t+\varphi)}+e^{-i (\omega_{D}(t) t+\varphi)})\\
\OmegaRabi(t) (e^{i (\omega_{D}(t)t+\varphi)}+e^{-i (\omega_{D}(t) t+\varphi)}) & -\omega_{0}
\end{pmatrix}.
\end{eqnarray}
The transformation  $\tilde{H} = U H_{L} U^{\dagger} + i \hbar \frac{\partial U}{\partial t} U^{\dagger}$ brings us into a reference frame rotating with the driving field frequency.
Here, $U$ is the unitary matrix $U = e^{-i/\hbar \int_0^T H_{D}(t) \textrm{d}t}$ defined by the applied driving field frequency:
\begin{eqnarray}
H_{D}(t) = \frac{\hbar}{2}
\begin{pmatrix}
\omega_{D}(t) & 0 \\
0 & -\omega_{D}(t)
\end{pmatrix}.
\end{eqnarray}
The time dependence of $\omega_{D}(t)$ causes the second term of the transformation $\left(i \hbar \frac{\partial U}{\partial t} U^{\dagger}\right)$ to be non-vanishing, which is in stark contrast to driving fields of fixed frequency.
Explicitly, for the Hamiltonian in the driving field frame (rotating at a time-varying rate) one obtains
\begin{eqnarray}
\tilde{H}(t) &=&
\frac{\hbar}{2}
\begin{pmatrix}
\Delta(t)+\dot{\Delta}(t)t & \OmegaRabi e^{i\omega_{D}(t)t} (e^{i (\omega_{D}(t)t+\varphi)}+e^{-i (\omega_{D}(t) t+\varphi)})\\
\OmegaRabi e^{-i\omega_{D}(t) t}(e^{i (\omega_{D}(t)t+\varphi)}+e^{-i (\omega_{D}(t) t+\varphi)})& -(\Delta(t)+\dot{\Delta}(t)t)
\end{pmatrix} \nonumber
\\
&=&
\label{eq:HamiltonianD}
\frac{\hbar}{2}
\begin{pmatrix}
\Delta(t)+\dot{\Delta}(t)t & \OmegaRabi e^{ i\varphi} \\
\OmegaRabi e^{- i \varphi} & -(\Delta(t)+\dot{\Delta}(t)t)
\end{pmatrix}.
\end{eqnarray}
where we rewrite the driving field detuning as $\Delta(t) = \omega_{0} - \omega_{D}(t)$ and its temporal derivative $\dot{\Delta}(t) = -\dot{\omega}_{D}(t)$.
In this derivation we performed the rotating wave approximation average out fast oscillating frequency components.

\section{Driving field parameters}
To ensure working in the experimentally accessible regime the free parameters were set to $\OmegaZ =\SI{3.5}{\MHz}$, $\DeltaZ=\SI{1}{\MHz}$ and $\tau = 2\pi\times 0.8/(2\OmegaZ)$ for our system limited by $\OmegaMax\approx \SI{7.5}{\MHz}$.
For completeness, find the expressions for the driving field parameters utilized to realize the superadiabatic geometric Pauli-X gate below.
\begin{eqnarray}
\OmegaRabi(t) = 
\left\lbrace
\begin{matrix}
\OmegaZ \left[1+\cos\frac{\pi t}{\tau} \right], & 0\leq t < \tau\\
\OmegaZ \left[1-\cos\frac{\pi (t - \tau)}{\tau} \right],& \tau \leq t  < 2 \tau\\
\OmegaZ \left[1+\cos\frac{\pi (t - 2\tau)}{\tau} \right], & 2 \tau \leq t < 3 \tau\\
\OmegaZ \left[1-\cos\frac{\pi (t - 3\tau)}{\tau} \right], & 3 \tau \leq t \leq 4 \tau 
\end{matrix}
\right.
\end{eqnarray}
and
\begin{eqnarray}
\DeltaS(t) = 
\left\lbrace
\begin{matrix}
\DeltaZ \left[\cos\frac{\pi t}{\tau} -1 \right], & 0\leq t < \tau\\
\DeltaZ \left[\cos\frac{\pi (t - \tau)}{\tau} +1 \right],& \tau \leq t  < 2 \tau\\
\DeltaZ \left[\cos\frac{\pi (t - 2\tau)}{\tau} -1 \right], & 2 \tau \leq t < 3 \tau\\
\DeltaZ \left[\cos\frac{\pi (t - 3\tau)}{\tau} +1 \right], & 3 \tau \leq t \leq 4 \tau 
\end{matrix}
\right.
\end{eqnarray}
and
\begin{eqnarray}
\varphi +\phi_{s}(t)
=
\left\lbrace
\begin{matrix}
\phi_{s}(t) + \widetilde{\varphi}_{1}', & 0\leq t < \tau \\
\phi_{s}(t) + \widetilde{\varphi}_{2}', & \tau \leq t < 3 \tau\\
\phi_{s}(t) + \widetilde{\varphi}_{1}', & 3\tau \leq t \leq 4\tau
\end{matrix}
\right.
.
\end{eqnarray}
As for the Pauli-Z gate the acquired geometric phase $\gamma'$ can be chosen via the relation $\gamma' = \pi - (\widetilde{\varphi}_{2}'-\widetilde{\varphi}_{1}')$.

In supplementary Supp.Fig.\ref{fig:sup:DrivingFieldParameter} the driving field parameter of the superadiabatic geometric Pauli-X and Pauli-Z gate used for the experimental realization are presented.
Note, while the Rabi-frequency $\OmegaS(t)$ and phase $\varphi + \phi_{s}(t)$ of the Pauli-X and Pauli-Z gate are identical except for a time shift of $\tau$, the driving field detuning functions $\Delta(t)$ defer strongly and the maximal absolute detuning of the Pauli-Z gate is twice as high as the one of the Pauli-X gate.
This difference is originating from the obtained solution $\Delta(t)$ of the differential equation $\Delta(t)+\dot{\Delta}(t)t=\DeltaS(t)$.
From the mathematical point of view the effective superadiabatic detuning $\DeltaS(t)$ are again identical up to a time shift of $\tau$.
However, involved with the different amplitudes of the detunings $\Delta(t)$ there might be some experimental implications for realizations choosing large $\DeltaZ$.
\begin{figure}[htbp]
\centering
\includegraphics[scale=1]{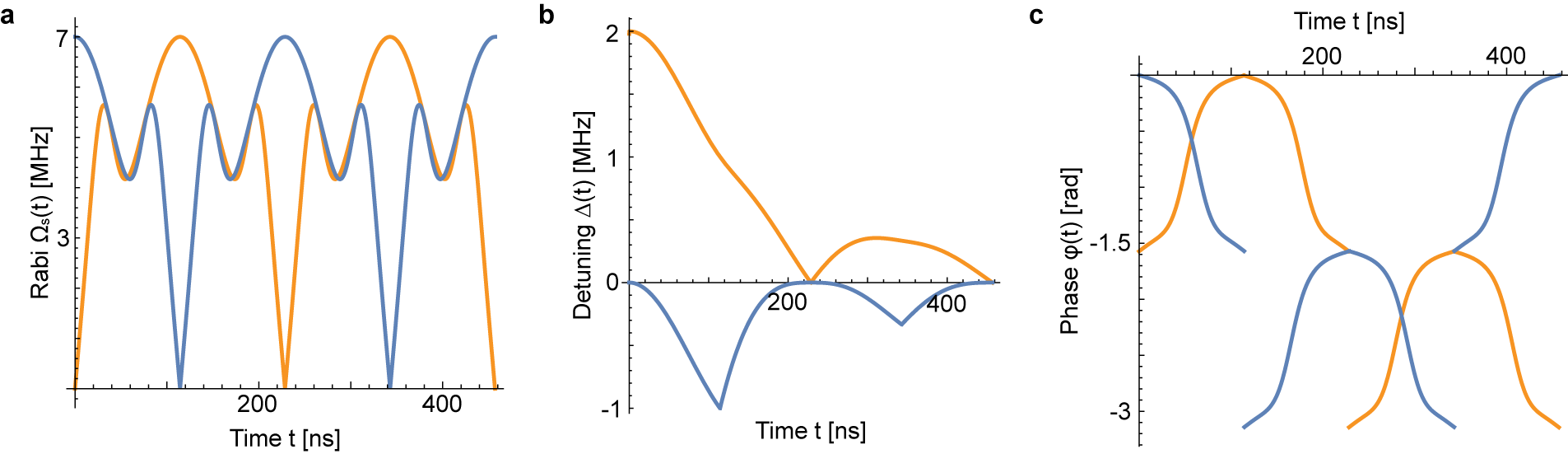}
\caption{
Driving field parameters as a function of time.
(a) Shows the applied Rabi frequency $\Omega_{\textrm{R}}(t)$ of the superadiabatic geometric quantum gate, (b) the driving field detuning $\Delta(t)$ (not to be confused with $\DeltaS(t)$) and (c) the phase $\varphi + \phi_{s}(t)$ for Pauli-X (blue) and Pauli-Z (orange) gate, respectively}
\label{fig:sup:DrivingFieldParameter}
\end{figure}

\section{Standard Quantum Process Tomography}

The standard quantum process tomography employed in this work is a modified version of the one presented in Ref.~\cite{2014ArroyoCamejoSupp}, which was designed for the reconstruction of the process matrix of evolutions in the three-dimensional Hilbert space $\mathcal{H}^3$.
Switching to two-dimensional $\mathcal{H}^2$ Hilbert space in the case of our super-adiabatic geometric realizations requires the choice of an appropriate set of basis operators.
We chose the Pauli matrices~$\bm{\sigma}$ complemented by the identity matrix (see Supp.Tab.\ref{tab:sup:BasisDefinition}).
The pulse sequences generating the basis states are illustrated in Supp. Tab.\ref{tab:sup:PulseSequence}.

\begin{table}[h]
\centering
\caption{Convention of the basis operators used in the quantum process tomography for the gates performed on the two dimensional Hilbert space $\mathcal{H}^2$.
}
\label{tab:sup:BasisDefinition}
\begin{tabular}{cccc}
$E_m$ & Pauli Operator & explicit expression & matrix representation \\ \hline \hline 
$E_1$ & $\sigma_{0}$ & $\Ketbra{0}{0}+\Ketbra{-}{-}$ &
$
\begin{pmatrix}
 1 & 0 \\
 0 & 1 
\end{pmatrix}
$
 \\
$E_2$ & $\sigma_{x}$ & $\Ketbra{0}{-}+\Ketbra{-}{0}$ &
$
\begin{pmatrix}
 0 & 1 \\
 1 & 0 
\end{pmatrix}
$ \\
$E_3$ & $\sigma_{y}$ & $i \Ketbra{0}{-}- i \Ketbra{-}{0}$ &
$
\begin{pmatrix}
 0 & -i \\
 i & 0 
\end{pmatrix}
$ \\
$E_4$ & $\sigma_{z}$ & $\Ketbra{0}{0}-\Ketbra{-}{-}$ &
$
\begin{pmatrix}
 1 & 0 \\
 0 & -1 
\end{pmatrix}
$ 
\end{tabular}
\end{table}

\begin{table}[h]
\centering
\caption{Pulse sequence for quantum process tomography. $\texttt{EXC}$ represent an initialization into the $m_{s} = 0$ state by a laser pulse.
$\left(\tau \right)_{j}$ symbolizes a microwave $j$-pulse of length $\tau$.
$\texttt{DET}$ means the readout out of the $m_{s} = 0$ population by excitation for $\SI{300}{\nano\second}$ and simultaneous fluorescence detection.} 
\label{tab:sup:PulseSequence}
\begin{tabular}{cccc}
$\Psi_{j}$ & explicit expression & initialization & readout \\ \hline \hline
$\Psi_{1}$ & $\Ket{0}$ & $\texttt{EXC}$ & $\texttt{DET}$ \\
$\Psi_{2}$ & $\Ket{-}$ & $\texttt{EXC} + \left(\pi\right)_{\overline{y}}$ & $\left(\pi\right)_{y} + \texttt{DET}$ \\
$\Psi_{3}$ & $\frac{1}{\sqrt{2}}\left(\Ket{0}+\Ket{-}\right)$ & $\texttt{EXC} + \left(\frac{\pi}{2}\right)_{\overline{y}}$ & $\left(\frac{\pi}{2}\right)_{y} + \texttt{DET}$\\
$\Psi_{4}$ & $\frac{1}{\sqrt{2}}\left(\Ket{0}+i\Ket{-}\right)$ & $\texttt{EXC} + \left(\frac{\pi}{2}\right)_{\overline{x}}$ & $\left(\frac{\pi}{2}\right)_{x} + \texttt{DET}$
\end{tabular}
\end{table}

\section{Randomized benchmarking}
The randomized benchmarking analysis is performed following the approach of Knill \textit{et al.}\cite{2008Knill}.
In essence, randomly generated sequences of gate operations of different length are utilized to measure the discrepancy between the expected and experimentally obtained result as a function of the sequence length.
In order to observe the error scaling independently of the applied sequence the errors are randomized by means of randomly chosen Pauli pulses which are interleaved with the $\pi/2$-rotations assumed to perform the computation.
The combination of one Pauli randomization pulse with one $\pi/2$-rotation is typically referred to as computational gate.
By additionally varying the sequence itself, an additional averaging is obtained.
The increase of the error probability with the number of computational gates $l$ leads to a decay of the average fidelity. The experimental randomized benchmarking data were fitted to the function $f(l) = 1-((1-\alpha_{n}\epsilon_{m})(1-\alpha_{n}\epsilon_{g})^{l} + 1)/\alpha_{n}$, where $\alpha_{n} = 2^n/(2^n-1)$ is a factor depending on the number of qubits $n$ involved\cite{2012Gaebler}. The error probability $\epsilon_{m}$ accounts for errors in the state preparation, the final projection Pauli randomization pulse combination and the readout.

The sequence generation is performed as follows:
$N_{G}=4$ random sequences $\mathcal{G}=\lbrace G_{1}, \dots \rbrace$ of $\pi/2$-rotations around the $x$, $z$, $\overline{x}$, and $\overline{z}$-axis are cropped to $N_{l}=13$ different lengths $l_{k}=\lbrace 2,4,6,8,10,14,18,22,26,30,34,40,48\rbrace$.
For each sequence the final state is calculated and a randomly chosen projective pulse $R$ added under the condition that the output state at the end of the sequence is one of the system eigenstates.
Each individual sequence is then Pauli randomized $N_{P}=8$ times by randomly choosing a sequence $\mathcal{P}=\lbrace P_{1},\dots \rbrace$ (consisting out of $\pi$-rotations around $x$, $z$, $y$, $\overline{x}$, $\overline{y}$, and $\overline{z}$-axis as well as the identity operation $\gamma=0,\pi$) of length $l_{k}+2$.
Subsequently, the total sequence $\mathcal{S} = P_{l_{k}+2}RP_{l_{k}+1}G_{l_{k}}\dots G_{1}P_{1}$ is generated.
By comparing the measured output state with the expected one the average gate fidelity is obtained.
The average gate fidelity is calculated as the mean over the $N = N_{P}\cdot N_{G}=32$ measurements for each gate length $l_{k}$.
Uncertainties are estimated by the standard error of the mean $\sigma_{\mathrm{mean}} = \sigma/\sqrt{N}$.
All operations in $\mathcal{S}$ are performed by SAGQGs.

Randomized benchmarking of the dynamic gate set follows equivalently when replacing the SAGQGs by dynamic $\pi$ and $\pi/2$-pulses.
The gates performing the computation are assumed to be $\pi/2$-rotations around $x$, $y$, $\overline{x}$, and $\overline{y}$-axis.
Rotations around the z-axis during the Pauli randomization are realized by appropriately adjusting the phase of the driving field.
Identity operations are performed by rotations of $2\pi$ around $x$ and $\overline{x}$.

\end{document}